\documentclass{article}
\usepackage{graphicx} 
\usepackage{amsmath,amssymb,exscale,graphicx,hyperref}
\usepackage{mathrsfs}
\usepackage{authblk}
\usepackage{tabularx}
\usepackage{multirow,enumitem}
\usepackage{slashed}
\usepackage{epsfig,psfrag,rotating,soul}
\usepackage{cite,color,url}
\usepackage[dvipsnames]{xcolor}
\usepackage{rotfloat}
\usepackage{hhline}
\usepackage{array}
\usepackage[normalem]{ulem}

\title{
Exploring enhanced non-resonant di-Higgs production at the HL-LHC with neural networks
}
\author[a]{Leandro~Da~Rold\thanks{daroldl@ib.edu.ar}}
\author[b]{Manuel~Epele\thanks{manuepele@fisica.unlp.edu.ar}}
\author[b]{Anibal~D.~Medina\thanks{anibal.medina@fisica.unlp.edu.ar}}
\author[b]{Nicol\'as~I.~Mileo\thanks{mileo@fisica.unlp.edu.ar}}
\author[b]{Alejandro Szynkman\thanks{szynkman@fisica.unlp.edu.ar}}
\affil[a]{Centro At\'omico Bariloche, Instituto Balseiro and CONICET, Av. Bustillo 9500, 8400, S. C. de Bariloche, Argentina}
\affil[b]{IFLP, CONICET - Dpto. de F\'{\i}sica, Universidad Nacional de La Plata, C.C. 67, 1900 La Plata, Argentina}

\begin{document}
\topmargin -1.0cm
\oddsidemargin -0.cm
\evensidemargin -0.cm

\maketitle

\begin{abstract}
We investigate di-Higgs production in the $b\bar{b}\gamma\gamma$ final state at the LHC, focusing on scenarios where the gluon fusion process is enhanced by new colored scalars, which could be identified as squarks or leptoquarks. We consider two benchmarks characterized by the mass of the lightest colored scalar, BM$_{\mathrm{L}}$ and BM$_{\mathrm{H}}$, corresponding to 464 GeV and 621 GeV, respectively. Using Monte Carlo simulations for both the signal and the dominant backgrounds, we perform a discovery analysis with deep neural networks, exploring various architectures and input variables. Our results show that the discrimination power is maximized by employing two dedicated classifiers, one trained against QCD backgrounds and another against backgrounds involving single-Higgs processes. Furthermore, we demonstrate that including high-level features -- such as the invariant masses $m_{\gamma\gamma}$, $m_{bb}$, and $m_{hh}$, as well as the transverse momenta and angular separations of the photon and $b$-jet pairs -- significantly improves the performance compared to using only low-level features as the invariant mass and momenta of the final particles. For the latter case, we find that architectures processing photon and $b$-jet variables separately can enhance the significance for BM$_{\mathrm{H}}$. Projecting for an integrated luminosity of 3 ab$^{-1}$, we obtain a significance of 7.2 for BM$_{\mathrm{L}}$, while it drops to 3.1 for BM$_{\mathrm{H}}$. In the particular case of BM$_{\mathrm{L}}$, discovery level significance can be reached at 1.4 ab$^{-1}$. {We also estimate how these results could be modified by potential systematic uncertainties.}
\end{abstract}

\newpage

\section{Introduction}

In 2012, the ATLAS and CMS Collaborations discovered a new particle consistent with the Standard Model (SM) Higgs boson, with a mass of approximately 125 GeV~\cite{ATLAS:2012yve,CMS:2012qbp}. This consistency was established through measurements of single Higgs production processes and detailed studies of Higgs properties, such as its spin, CP nature, and decays to Standard Model particles.
The study of di-Higgs and tri-Higgs production is sensitive to the Higgs cubic and quartic self-interactions, providing direct information about the Higgs potential and the mechanism of electroweak symmetry breaking~\cite{Dicus:1987ic,Glover:1987nx,Plehn:1996wb,Dawson:1998py,Djouadi:1999rca,BarrientosBendezu:2001di}. Measuring di-Higgs production at the Large Hadron Collider (LHC) is one of the main goals of the high luminosity LHC (HL-LHC), though particularly challenging for a SM-like Higgs. 
Currently, with an integrated luminosity of 138 fb$^{-1}$, CMS has established an upper limit of 3.5 times the Standard Model cross-section for inclusive di-Higgs production~\cite{CMS:2025ngq}.
Studies based on the $\mathcal{L}=140$~fb$^{-1}$ of integrated luminosity dataset collected at $\sqrt{s}=13$~TeV indicate that, extrapolating to $\mathcal{L}=3$~ab$^{-1}$ at $\sqrt{s}=14$~TeV (HL-LHC), the expected significance for observing this process would be $\sim 7 \sigma$ in the context of the SM~\cite{CMS:2025hfp}.

There is considerable interest in physics beyond the Standard Model (BSM) that could enhance the di-Higgs production cross-section, thereby increasing the significance or enabling the study of this process with luminosities smaller than 3~ab$^{-1}$. An example of a minimal BSM scenario that has received considerable attention is via  modifications of the Higgs quartic coupling~(see for example \cite{ATLAS:2017muo,Barr:2014sga,Kling:2016lay,Adhikary:2017jtu,Goncalves:2018qas,Belfkir:2025zlx}) which could increase di-Higgs production. Another enhancement can occur in the di-Higgs gluon-fusion production channel,  where there can be new contributions from colored particles in the loop such as is the case in some supersymmetric models~\cite{Plehn:1996wb,Belyaev:1999mx}, in theories with vector-like fermions~\cite{Dib:2005re,Dawson:2012mk,Gillioz:2012se} or in models with new scalar fields~\cite{Enkhbat:2013oba,Huang:2017nnw,Bhaskar:2022ygp,DaRold:2021pgn}, among other possibilities~\cite{Cao:2013si,Hespel:2014sla,Dawson:2015oha,Batell:2015koa}. The enhancement should be achieved keeping the single Higgs observables under control~\cite{Crivellin:2020ukd}. In Ref.~\cite{DaRold:2023hst} we studied this production channel in a framework with new colored scalar fields, which can be identified either with the quark scalar superpartners or with a set of scalar leptoquarks. Focusing on the $b\bar b\gamma\gamma$ final state, we analyzed the impact of the new states on differential kinematical distributions. We also studied the luminosities required for discovery and exclusion in scenarios characterized by the mass of the lightest new state. The analysis was performed by applying optimized cuts on kinematical variables.

In this work, we study once again di-Higgs production at the LHC in the $b\bar b\gamma\gamma$ final state, within the same BSM model considered in Refs.~\cite{DaRold:2021pgn,DaRold:2023hst}, but we shift the strategy and consider using deep neural network (DNN) classifiers. We generate both the signal and the main backgrounds through Monte Carlo simulations, and train DNNs to distinguish between them. We explore various DNN architectures and different sets of kinematical input variables. We find that training two separate classifiers, one against the QCD-induced backgrounds and another against backgrounds that involve the production of one Higgs, improves the discrimination power, with the invariant masses $m_{\gamma\gamma}$, $m_{bb}$, and $m_{hh}$ being effective discriminating observables. We also find that, in some cases, architectures with dedicated sub-networks that separately process the photon and $b$-jet variables can lead to an improved significance.

The article is organized as follows: in Sec.~\ref{sec:di-Higgs-prod} we present the model with new colored scalars and introduce two benchmark scenarios. We also briefly discuss the BSM contributions to di-Higgs production at the LHC, together with the main backgrounds. In Sec.~\ref{sec:sim}, we describe the tools used for the simulations and define a set of basic cuts. Sec.~\ref{sec:NNs} presents the main analysis and results: we describe the DNN architectures employed, show the classifier performance, and compute the expected significances. Finally, conclusions are given in Sec.~\ref{sec:conclusions}.

\section{Di-Higgs production with colored scalars}
\label{sec:di-Higgs-prod}

We consider a model with colored scalar fields in the same representations as one generation of quarks: $\phi_2\sim({\bf 3},{\bf 2})_{1/6}$, $\phi_1\sim(\bar{\bf 3},{\bf 1})_{1/3}$ and $\bar \phi_1\sim(\bar{\bf 3},{\bf 1})_{-2/3}$, that can be associated with either supersymmetric partners of one generation of quarks or with a set of leptoquarks. The full Lagrangian, as well as its phenomenology, was presented in Ref.~\cite{DaRold:2021pgn} for the case of leptoquarks. The model includes four mass eigenstates, two up type and two down type, respectively of charge $2/3$ and $-1/3$. The potential in terms of these states can be written as
\begin{equation}
V=\sum_{i,q}|\chi^q_i|^2m_{\chi^q_i}^2 + h \sum_{i,j,q}\chi^{q*}_i\chi^q_j \mathscr{C}^{(q)}_{ij} + h^2 \sum_{i,j,q}\chi^{q*}_i\chi^q_j {\cal Q}^{(q)}_{ij} + \dots
\label{eq-Vmass}
\end{equation}
where $i,j=1,2$ and $q=u,d$. $\mathscr{C}^{(q)}$ and ${\cal Q}^{(q)}$ are the cubic and quartic couplings in the mass basis.

We choose two benchmark (BM) scenarios: BM$_{\mathrm{L}}$ and BM$_{\mathrm{H}}$, where the mass of the lightest state is $m_{\rm NP}=464$~GeV and 621~GeV, respectively (for further details on the rest of the spectrum and couplings, see Ref.~\cite{DaRold:2021pgn}).  
It is important to emphasize that none of the chosen BM points are ruled out by experimental constraints from electroweak precision observables, flavor physics or collider searches, in particular from single Higgs and QCD-induced direct leptoquark productions. These constraints were studied in detail in Ref.~\cite{DaRold:2021pgn} where we refer the interested reader. No new updated versions of the experimental searches these constraints are based on were conducted since then at the moment and thus the benchmarks BM$_{\mathrm{L}}$ and BM$_{\mathrm{H}}$ are currently not ruled out experimentally.

\begin{figure}[h!]
        \centering
    \includegraphics[width=0.35\textwidth]{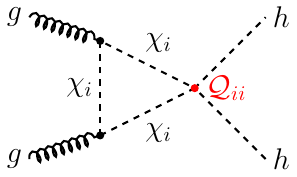}\hspace{2cm}
    \includegraphics[width=0.35\textwidth]{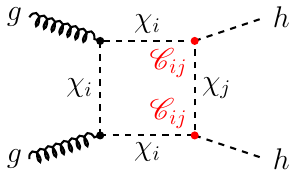}
    \caption{Representative Feynman diagrams for di-Higgs production from gluon fusion with colored scalar fields.}
        \label{fig-1}
\end{figure}

The di-Higgs production cross-sections from gluon fusion in this model were calculated in Refs.~\cite{DaRold:2021pgn,DaRold:2023hst} (see also Ref.~\cite{Enkhbat:2013oba}). 
In Fig.~\ref{fig-1} we show two representative Feynman diagrams for this process.
By using the PDF set PDF4LHC15\_nnlo\_mc~\cite{Butterworth_2016} and {\tt LoopTools}~\cite{Hahn:1998yk} the LO cross-section for the two benchmark points and for the SM were computed, obtaining the results of Table~\ref{t-xs}. 
\begin{table}[h]
\renewcommand*{\arraystretch}{1.5}
\begin{center}
\begin{tabular}{|c|c|c|}
\hline
$\sigma_{hh}$ & BM$_{\mathrm{L}}$ & BM$_{\mathrm{H}}$ 
\\ \hhline{|=|=|=|}
$(\sigma_{hh}/\sigma^{\rm SM}_{hh})^{\rm LO}$ & 2.16  & 1.37 \\ 
\hline
\end{tabular}
\caption{Di-Higgs production cross-section for the BMs, normalized with respect to the SM one at LO.}
\label{t-xs}
\end{center}
\end{table}

{To estimate the corrections to the cross-section, we used the NNLL SM result for $m_h=125$~GeV reported in Ref.~\cite{LHCHiggsCrossSectionWorkingGroup:2016ypw}, $\sigma^{\rm NNLL}=39.59$~fb, and worked under the assumption that the leading higher-order corrections\footnote{For computations in the SM see Refs.~\cite{deFlorian:2013uza,deFlorian:2013jea,Maltoni:2014eza,Frederix:2014hta,Grigo:2014jma,Grigo:2015dia,Borowka:2016ehy,Degrassi:2016vss,deFlorian:2016uhr,Spira:2016zna,Borowka:2016ypz,Heinrich:2017kxx,Jones:2017giv,Banerjee:2018lfq,Baglio:2018lrj,Chen:2019lzz,Chen:2019fhs,Baglio:2020ini,Muhlleitner:2022ijf,AH:2022elh,Bi:2023bnq}.} cancel in the ratio. This approximation is partially supported by Fig.~2 in Ref.~\cite{Agostini:2016vze}, which shows that the NLO K-factor for an analogous setup with squarks in the MSSM is similar to the SM one. It should be noted, however, that the results presented in Ref.~\cite{Agostini:2016vze} are based on calculations performed in the limit of vanishing external momenta. Although this approximation is known to be less reliable for double-Higgs production than for single-Higgs, it remains the best estimate available, given that no higher-order QCD corrections have been computed for this specific model\footnote{In principle, a differential K-factor would reweight the kinematical distributions, not only the overall normalization, and could modify the analysis. However, such a differential K-factor has not been computed for this specific model.}.}

\begin{figure}[h!]
    \centering
    \includegraphics[width=0.45\textwidth]{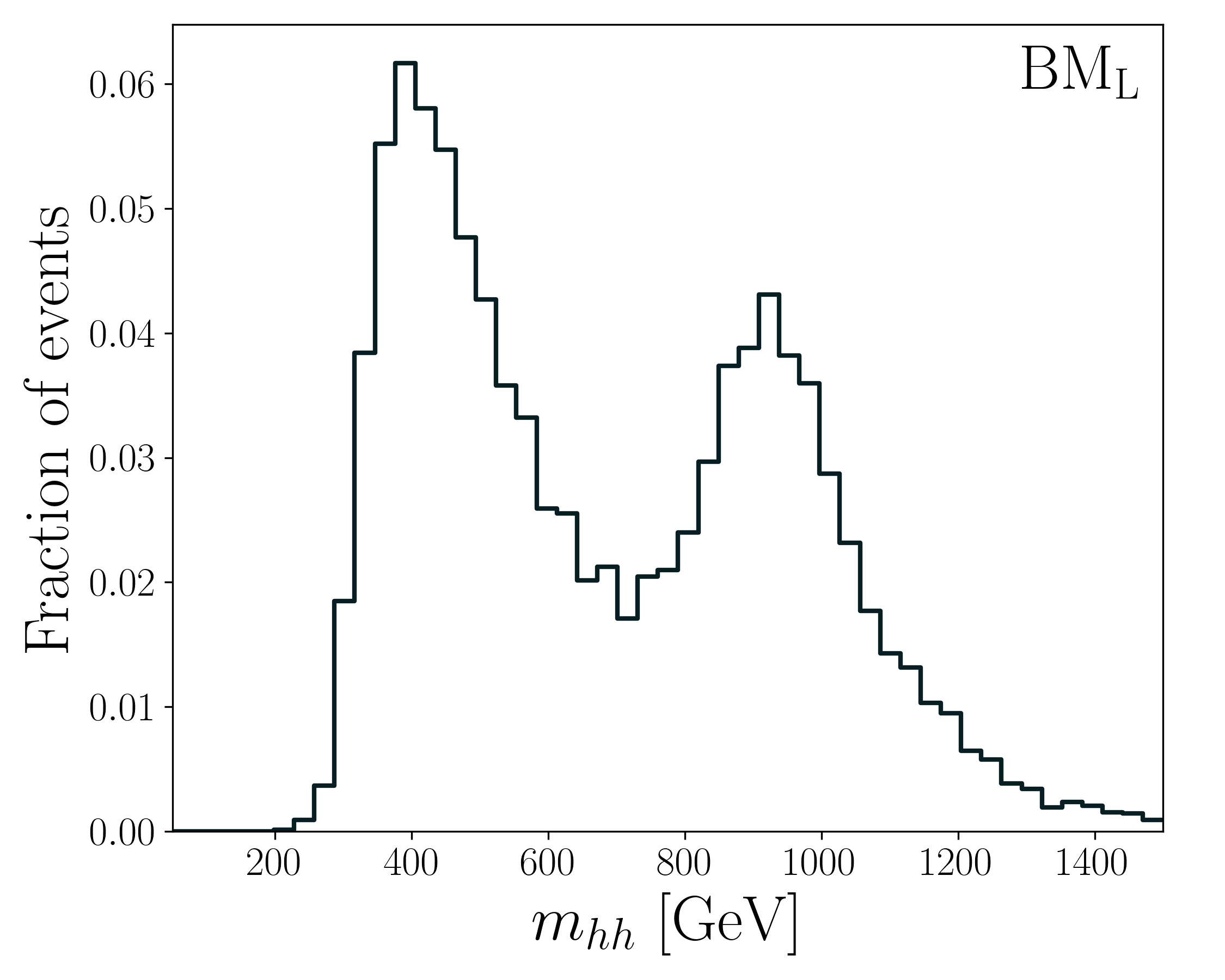}
    \includegraphics[width=0.45\textwidth]{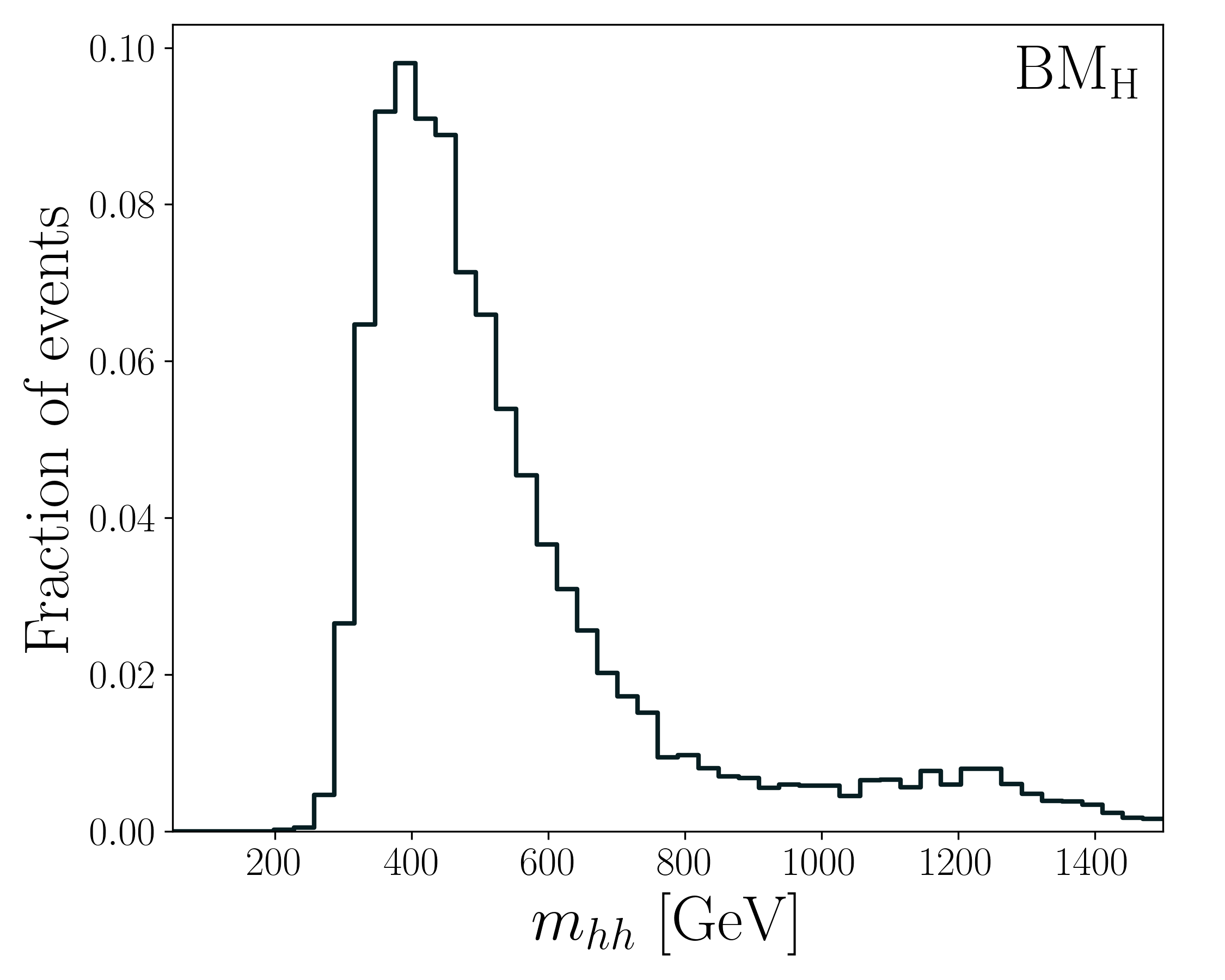}
    \caption{Normalized distributions of the di-Higgs invariant mass $m_{hh}$ for the two BMs considered in this work at detector level and after applying a set of selection requirements, see section \ref{sec:sim}.}
    \label{fig-2}
\end{figure}

The effects of the new colored scalar states in the differential distributions have been analyzed in Ref.~\cite{DaRold:2023hst}. In the invariant mass and in the transverse momentum distributions, $m_{hh}$ and $p_{Th}$ respectively, the top peak is enhanced and a new peak around $m_{hh}\simeq 2 \, m_{\rm NP}$ and $p_{Th}\simeq m_{\rm NP}$ appears, signaling the presence of the lightest new state, see Figs. 2 and 3 of Ref.~\cite{DaRold:2023hst}. As a consequence, the Higgs bosons can be boosted and their decay products are collimated, populating the region of small $\Delta R_{\gamma\gamma}$ and $\Delta R_{bb}$. These effects are diluted as $m_{\rm NP}$ increases. We have updated the $m_{hh}$ distribution at detector level, increasing the statistics with respect to~\cite{DaRold:2023hst}, see Fig.~\ref{fig-2}. Note that for BM$_{\mathrm{L}}$ the new peak in the $m_{hh}$ distribution is of order $60\%$ of the top peak, whereas for BM$_{\mathrm{H}}$ it is only $10\%$.

It is clear from the previous discussion that the selected benchmarks cover two complementary scenarios: BM$_{\mathrm{L}}$ looks more promising to discriminate against backgrounds, while BM$_{\mathrm{H}}$ seems more challenging.

Due to their large cross-sections, the dominant backgrounds are the QCD-induced processes $b\bar b\gamma\gamma$, $b\bar b\gamma j$ and $c\bar c\gamma\gamma$, with a jet misidentified as a $\gamma$ and $c$-jets faking $b$-jets. In particular, we considered a misidentification of $j$ as $\gamma$ of $0.5\times 10^{-3}$, whereas for the misidentification of $c$-jets as $b$-jets we used the rate in the default HL-LHC card for the ATLAS detector provided by Delphes. In addition, though their cross-sections are smaller, we took into account the single-Higgs processes $Zh$ and $t \bar t h$ (dubbed Single-$h$ from now on), with the pair of photons coming from the decay of the Higgs boson, making them similar to the signal.

\section{Simulations}
\label{sec:sim}
Signal and backgrounds events are simulated with {\tt MadGraph5\_aMC@NLO} (MG5) \cite{Alwall:2011uj,Alwall:2014hca} at $\sqrt{s}=14$~TeV with the PDF set PDF4LHC15\_nnlo\_mc including one extra jet with the MLM scheme for the jet-parton matching~\cite{Mangano:2006rw}. The Higgs decay is simulated with {\tt MadSpin}~\cite{Artoisenet:2012st}, the parton showering and hadronization is handled with {\tt Pythia~8}~\cite{Sjostrand:2006za,Sjostrand:2007gs,Sjostrand:2014zea} and a fast detector simulation is carried out with {\tt Delphes 3}~\cite{deFavereau:2013fsa} using the ATLAS HL-LHC card.
 
In order to simulate the signal, which corresponds to a process at one loop, we implement the model with the colored scalars fields in {\tt FeynRules}~\cite{Degrande:2011ua,Alloul:2013bka} at tree level and generate a Lagrangian renormalized at one loop by making use of {\tt FeynArts}~\cite{Hahn:2000kx} and {\tt NLOCT}~\cite{Degrande:2014vpa}. The resulting NLO model is exported in UFO format~\cite{Darme:2023jdn} to be used for simulations in MG5 and is available in \url{https://github.com/manuepele/SM_LQs.git}. In order to validate the UFO model, we have obtained the di-Higgs production cross-section at LO as well as the $m_{hh}$ and $p_{Th}$ distributions from parton level events generated with MG5 and checked that they are consistent with our previous computation of the same quantities based on {\tt LoopTools}.

We apply the following set of requirements at detector level~\cite{ATLAS:2017muo}:
\begin{itemize}
\item At least two isolated photons with $p_T >$ 30 GeV and $|\eta|<1.37$ or $1.52<|\eta|<2.37$.
\item At least two $b$-tagged jets with $|\eta|< 2.4$ and $p_T > 40$ GeV and 30 GeV, for the leading and subleading $b$-jets, respectively.
\item From all the possible candidate pairs of photons and $b$-jets we select those with the invariant mass closest to the Higgs mass and, for those pairs, we require $p^{\gamma\gamma}_T, p^{bb}_T> 80$ GeV, $0.4<\Delta R_{\gamma\gamma}<2.0$, $0.4 < \Delta R_{bb}<2.0$ and $\Delta R_{\gamma j}>0.4$. 
\item Events with more than five jets with $|p_T|>30$ GeV and $|\eta|<2.5$ (including the tagged $b$-jets with $|\eta|<2.4$) or with isolated leptons with $p_T> 25$ GeV and $|\eta|<2.5$ are vetoed.
\item We also included the requirements $60 \text{ GeV }< m_{\gamma\gamma} < 200\text{ GeV }$ and $m_{bb} < 45\text{ GeV }$.
\end{itemize}
In the following, the acceptance corresponding to this set of cuts is denoted as $\mathcal{A}$.

\section{Machine Learning Analysis}
\label{sec:NNs}
The classifiers used in this work are built by training DNNs of multilayer perceptron type (MLP), with the input layer consisting of a set of kinematic variables associated to the objects reconstructed in the final state. In particular, for our analysis, we use the pair of photons and $b$-jets passing the acceptance requirements described in Sec.~\ref{sec:sim}. The features considered in this study are listed in Table~\ref{tab:vars} along with a brief description, where we split them in low-level features (LLF), that correspond to $p_T,\,\eta$ and $\phi$ of the four objects in the final state (for the $b$-jets we add the invariant mass of each of them), and high-level features (HLF), which are functions of the LLF.

We split the simulated sample of events surviving the acceptance requirements into training and testing subsamples. The training subsample contains an equal proportion of signal and background events; the 20\% of the training events is used for validation. For the testing subsample, we follow the criterion of including at least 10 times the expected number of events at $\mathcal{L}=3$ ab$^{-1}$ of each process (signal and backgrounds), based on their cross-sections and selection acceptances. 

All events are preprocessed to normalize the range of variation of every input feature in such a way that the training set has zero mean and unit variance. To control the training procedure and prevent overfitting, we used the early-stopping technique with a 50-epoch tolerance based on the area under the curve metric (AUC) over the validation. The training stage is driven by the Adam optimization algorithm, with a global learning rate of $0.001$. We also implemented a learning rate scheduler algorithm to optimize the free parameters of the model in the final stage of the training, which reduces the learning rate by one order of magnitude if the AUC cannot be improved after $40$ epochs.

DNNs fitted via stochastic algorithms can be sensitive to specific patterns in the training sample and may find a different set of weights each time they are trained. We handle model fluctuations by training ensembles of 80 models for each BM. The models are finally combined by averaging the model output for every event over the ensemble.

\begin{table}[h]
    \centering
    \renewcommand{\arraystretch}{1.5}
    \begin{tabular}{c|c|c}
    \multirow{5}{*}{Low-level Features} & Name     & Description \\
    \hline
     & $p^j_T$ & Transverse momentum of $j=\gamma_1,\gamma_2,b_1,b_2$ \\
    & $\eta_j$ & Pseudorapidity of $j=\gamma_1,\gamma_2,b_1,b_2$ \\
    & $\phi_j$ & Azimuthal angle of $j=\gamma_1,\gamma_2,b_1,b_2$ \\
    & $m_j$ & Invariant mass of $j=b_1,b_2$ \\[1mm]
    \hline
    \multirow{7}{*}{High-level Features} & $m_{\gamma\gamma}$ & Invariant mass of the photon pair \\
    & $m_{bb}$ & Invariant mass of the $b$-jet pair \\
    & $m_{hh}$ & Invariant mass of the reconstructed Higgs boson pair \\
    & $\Delta R_{ij}$ & Distance $\sqrt{\Delta\eta_{ij}^2+\Delta\phi_{ij}^2}$, $i,j=\gamma_1,\gamma_2,b_1,b_2$ \\
    & $p^{\gamma\gamma}_T$ & Transverse momentum of the photon pair \\
    & $p^{bb}_T$ & Transverse momentum of the $b$-jet pair \\
    & MET & Missing transverse energy \\
    \hline
    \end{tabular}
    \caption{List of the features used in this study. The subscripts 1 and 2 denote the leading and subleading object ($\gamma$ or $b$-jet) ordered by $p_T$. }
    \label{tab:vars}
\end{table}

\subsection{Analysis with all features}
\begin{figure}[h!]
    \centering
    \includegraphics[width=.65\linewidth]{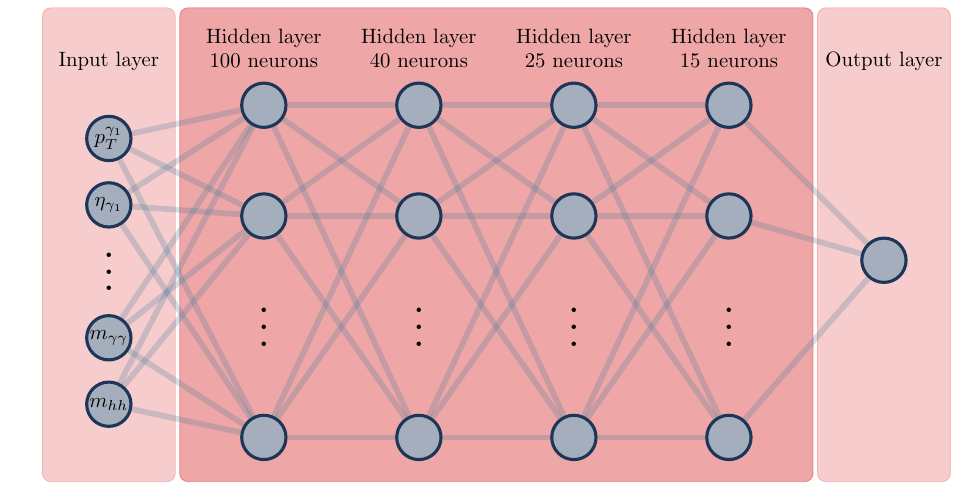}
    \caption{DNN architecture used for the analysis with all the features.}
    \label{fig:arch_A}
\end{figure}

Since we are interested in maximizing the signal discovery significance at the LHC, we first consider in our study all features (low and high) shown in Table~\ref{tab:vars}. It is clear then that by taking into account the high-level features we are providing the DNNs with some physics knowledge of the problem to be analyzed. Throughout this section we consider classifiers based on the architecture shown in Fig.~\ref{fig:arch_A}. As a first step, we train the signal against all the backgrounds, each weighted by the product of the cross-section and the selection acceptance, finding that the resulting classifier is able to separate successfully the signal from the QCD backgrounds but not from the Single-$h$ contribution. This limitation is expected since QCD processes have much larger cross-sections, dominating the training procedure. In order to achieve a better discrimination of the Single-$h$ backgrounds, we combine two classifiers: one trained exclusively with QCD backgrounds, and the other with the Single-$h$ backgrounds. 

\begin{figure}[h!]
    \centering
    \includegraphics[width=0.455\linewidth]{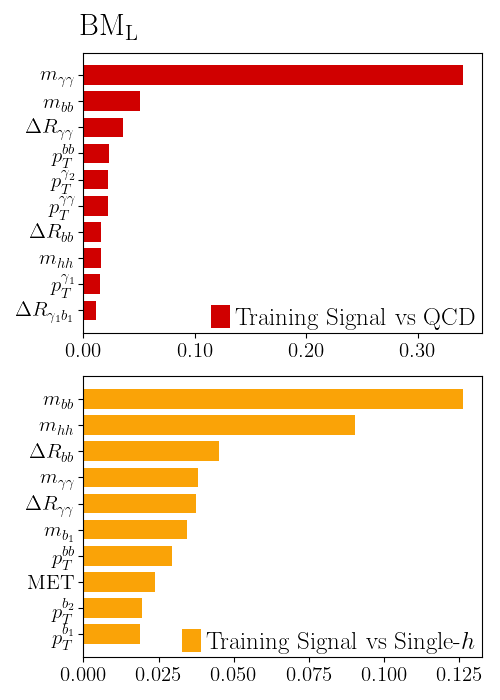} 
    \includegraphics[width=0.455\linewidth]{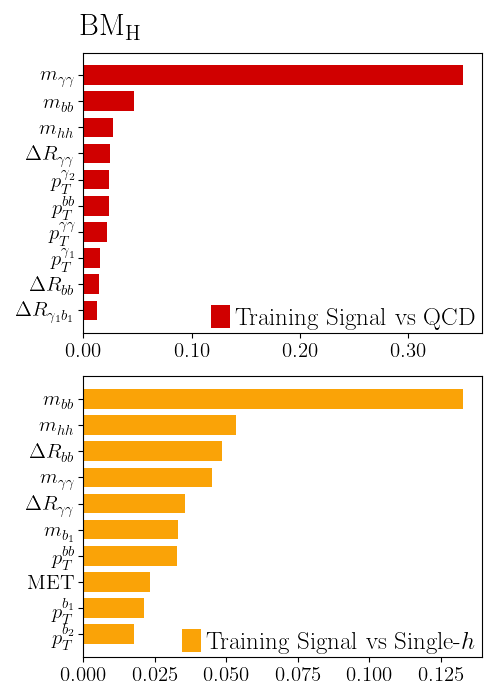}
    \caption{Ranking of features based on the Shapley values approach for classifiers trained against the QCD (upper panels) or Single-$h$ backgrounds (lower panels).}
    \label{fig:ranking}
\end{figure}

In Fig.~\ref{fig:ranking} we show the ranking, based on the Shapley values, of the ten most important variables used as input for the two classifiers during the training. For the one dedicated to separate the QCD backgrounds, $m_{\gamma\gamma}$ is the most important feature, while for the classifier dedicated to Single-$h$, it is $m_{bb}$, with $m_{\gamma\gamma}$ dropping to the fourth place since photons originate from the Higgs decay as happens with the signal. In the case of the classifier dedicated to separate the signal from the QCD backgrounds, one may expect $m_{bb}$ to be ranked closer to $m_{\gamma\gamma}$ since in principle both variables are very efficient to discriminate the signal from the QCD background. However, showering, hadronization and detector effects lead to a broadening of the resonances such that, in particular for $m_{bb}$, the resonant peak shifts to lower masses and flattens, extending several tens of GeV at either side, which degrades its discrimination power considerably~\cite{ATLAS:2017muo,CMS:2020tkr,ATLAS:2025hhd}. Although there are variables such as $m_{\gamma\gamma}$ or $m_{bb}$ that are highly powerful at discriminating the signal from the backgrounds, the rest of the features cannot be excluded from the analysis given that the initial signal-to-background ratio is very small. This makes other lower-ranked variables also relevant in maximizing the performance of the classifiers.

\begin{figure}[h!]
    \centering
    \includegraphics[width=.95\linewidth]{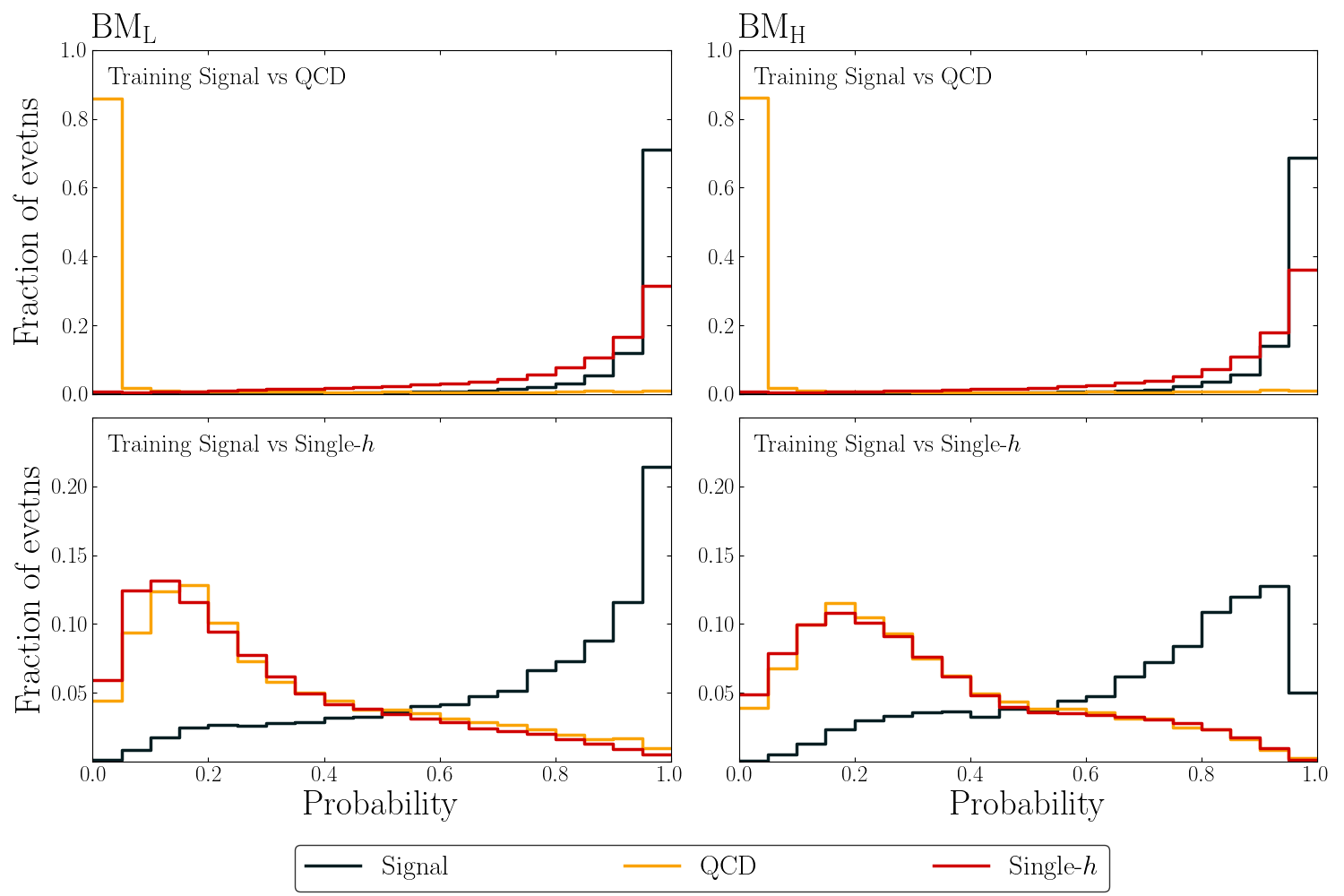}
    \caption{Probability distributions given by the output of DNN classifiers trained with background events corresponding only to QCD (upper panels) or Single-$h$ (lower panels) processes and signal events of BM$_{\mathrm{L}}$ (left panels) or BM$_{\mathrm{H}}$ (right panels).}
    \label{fig:trainings}
\end{figure}

The output distributions of the two classifiers for BM$_\mathrm{L}$ and BM$_\mathrm{H}$ testing samples are shown in Fig.~\ref{fig:trainings}~\footnote{When interpreting the plots of Fig.~\ref{fig:trainings} one should take into account that they are normalized to unity for better visualization.}. From the top panels, we see that the classifier dedicated to QCD backgrounds fails to discriminate the Single-$h$ backgrounds from the signal. On the other hand, as can be seen from the lower panels, the classifier dedicated to Single-$h$ is able to separate both backgrounds from the signal with almost the same performance, although its discrimination power of the QCD backgrounds is substantially worse than the dedicated classifier. The fact that the classifier dedicated to the Single-$h$ backgrounds is able to discriminate also the QCD contribution, even when no events from it are included in the training sample, can be understood by looking at the ranking of inputs. The most important variable for the classifier dedicated to QCD is $m_{\gamma\gamma}$, which is not particularly discriminant for the Single-$h$ backgrounds where the pair of photons reconstructs the Higgs mass. For the classifier dedicated to Single-$h$, in contrast, the most important variable is $m_{bb}$, which is useful to separate both backgrounds from the signal. Furthermore, given that the shapes of the red and yellow histograms in the bottom panels are quite similar, one can conclude that 
the chosen set of variables used for the classification is enough to achieve the maximum discrimination between the signal and QCD backgrounds, when the classifier is trained only with the Single-$h$ backgrounds.~\footnote{In fact, if one removes $m_{\gamma\gamma}$, the QCD distribution shifts to the region of larger output values, while the Single-$h$ distribution does not change.}

Once the two classifiers have been trained, they are used on the testing set assigning an output per event for both of them. As a rule, a given event is labeled as signal if both outputs are above thresholds set for each classifier, which are chosen to optimize the discovery significance (see Table~\ref{tab:algo}). Such significance is computed as
\begin{equation}
\label{eq:sig}
    \mathcal{S}=\sqrt{-2\left((s+b)\ln\left(\frac{b}{s+b}\right)+s\right)},
\end{equation}
where $s = \mathcal{L}\times \sigma_s \times \mathcal{A}_s \times \text{TPR} $ and $b = \sum_i b_i$, with $b_i = \mathcal{L}\times \sigma_i \times \mathcal{A}_i \times \text{FPR}_i$ and $i$ labeling the backgrounds. $\text{TPR}$ and $\text{FPR}$ stand for the true and false positive rates, respectively. From now on, we refer to the combination of both classifiers as All Features (AF) classifier.

\renewcommand{\tabularxcolumn}[1]{m{#1}}
\begin{table}[ht]
\centering
\small
\renewcommand{\arraystretch}{2.4}
\begin{tabularx}{\textwidth}{ 
    >{\centering\arraybackslash}m{3.5cm}
     >{\raggedright\arraybackslash}X             
}
\hline
\multicolumn{2}{l}{\normalsize \textbf{Algorithm for the implementation of the AF classifier}} \\ \hline
\textbf{Step} & {\hspace{5cm}\textbf{Description}} \\ \hline
1. Training & 
a) Train a classifier exclusively with signal vs. QCD backgrounds. \newline
b) Train a classifier exclusively with signal vs. Single-$h$ backgrounds. \\ 

2. Evaluation & Apply both classifiers to the testing set, assigning two output scores ($O_{\text{QCD}}, O_{\text{Single-}h}$) to each event. \\ 

3. Labeling & Label an event as signal if it satisfies the condition: 
\begin{equation*}
(O_{\text{QCD}} > \mathcal{E}_{\text{QCD}}) \land (O_{\text{Single-}h} > \mathcal{E}_{\text{Single-}h})
\end{equation*}
where $\mathcal{E}_{\text{QCD}}$ and $\mathcal{E}_{\text{Single-}h}$ are the respective classifier thresholds. \\ 

4. Optimization & Perform a two-dimensional scan over the $(\mathcal{E}_{\text{QCD}}, \mathcal{E}_{\text{Single-}h})$ parameter space to find the threshold values that maximize the discovery significance $\mathcal{S}$ (Eq.~\ref{eq:sig}). \\ \hline
\end{tabularx}
\caption{Description of the algorithm applied to implement the AF classifier for signal vs. background discrimination.}
\label{tab:algo}
\end{table}

\begin{figure}[h!]
    \centering
    \includegraphics[width=0.45\linewidth]{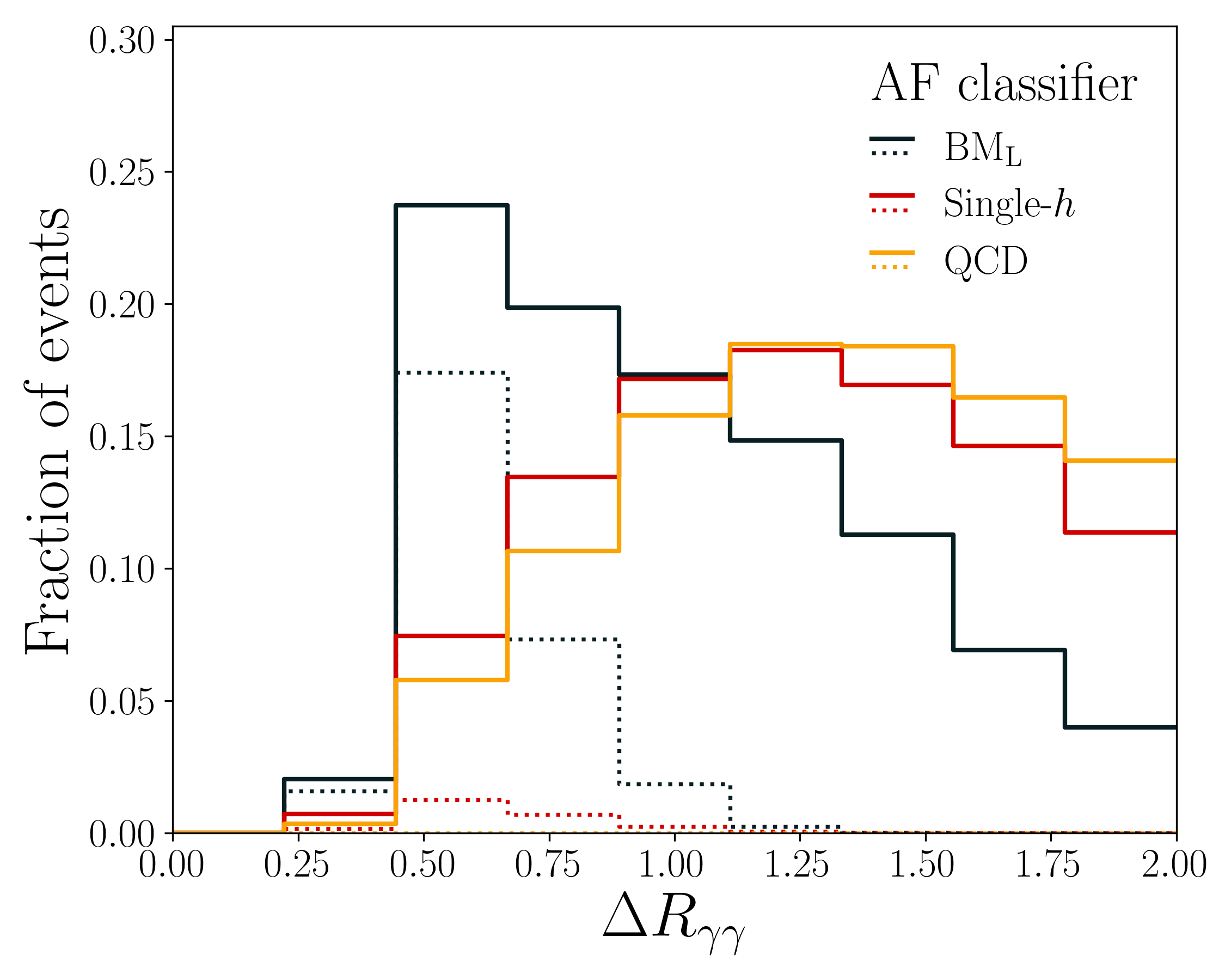}
    \includegraphics[width=0.45\linewidth]{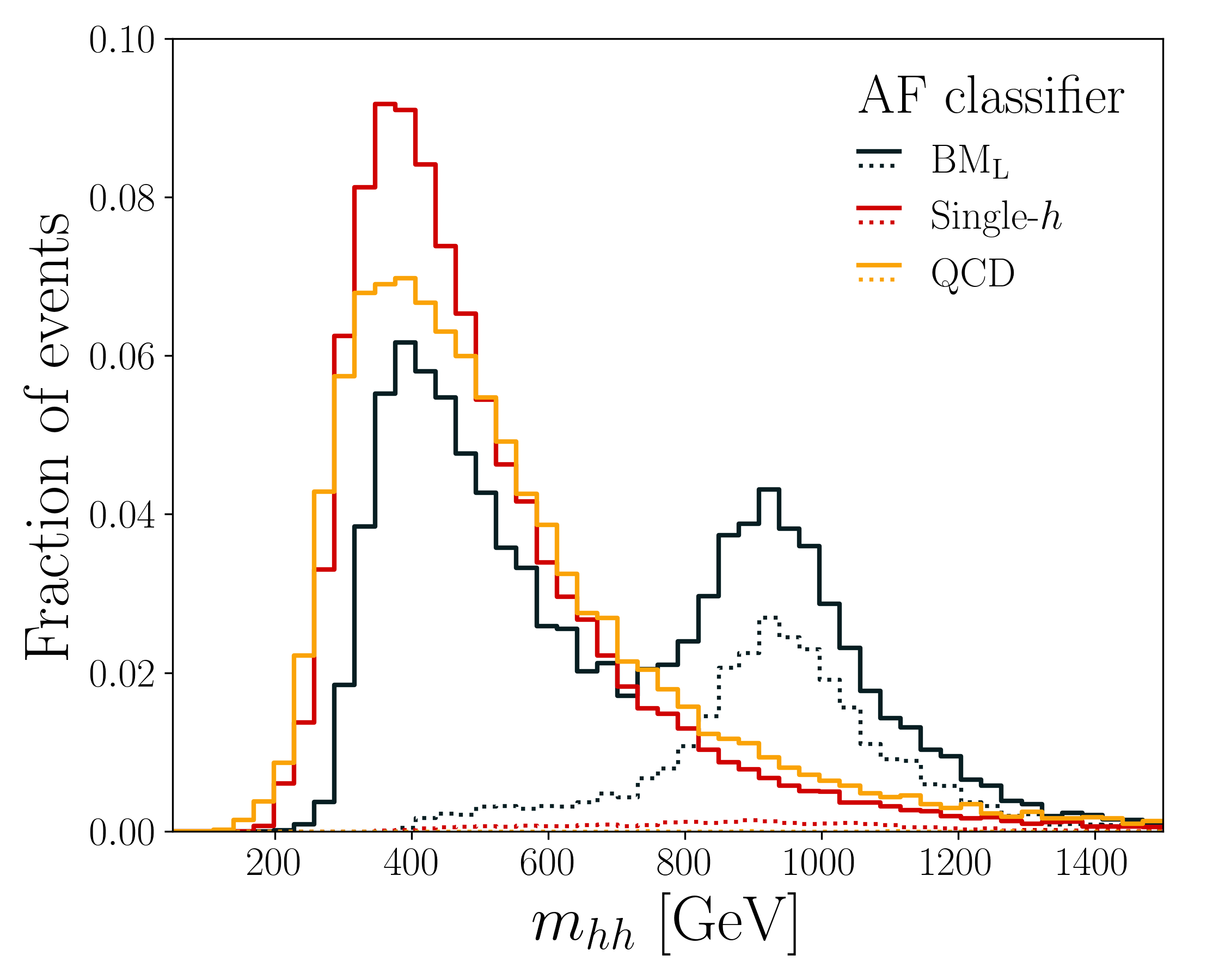}
    \caption{$\Delta R_{\gamma\gamma}$ (left panel) and $m_{hh}$ (right panel) distributions for BM$_{\mathrm{L}}$, before (solid lines) and after (dotted lines) the classification with the AF classifier (the distribution of misclassified QCD background is not visible in the figure).}
    \label{fig:distBM1}
\end{figure}

\begin{figure}[h!]
    \centering
    \includegraphics[width=0.45\linewidth]{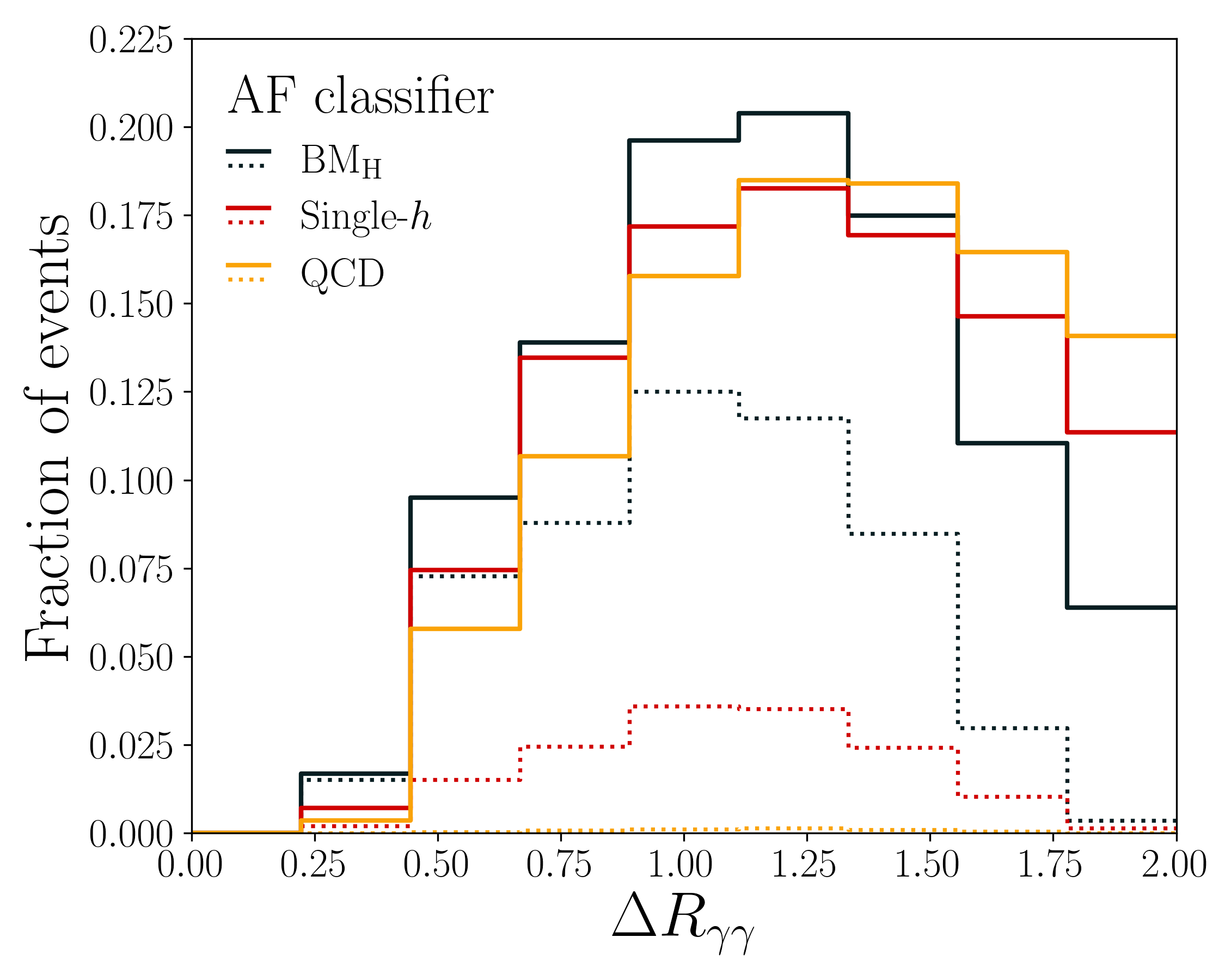}
    \includegraphics[width=0.45\linewidth]{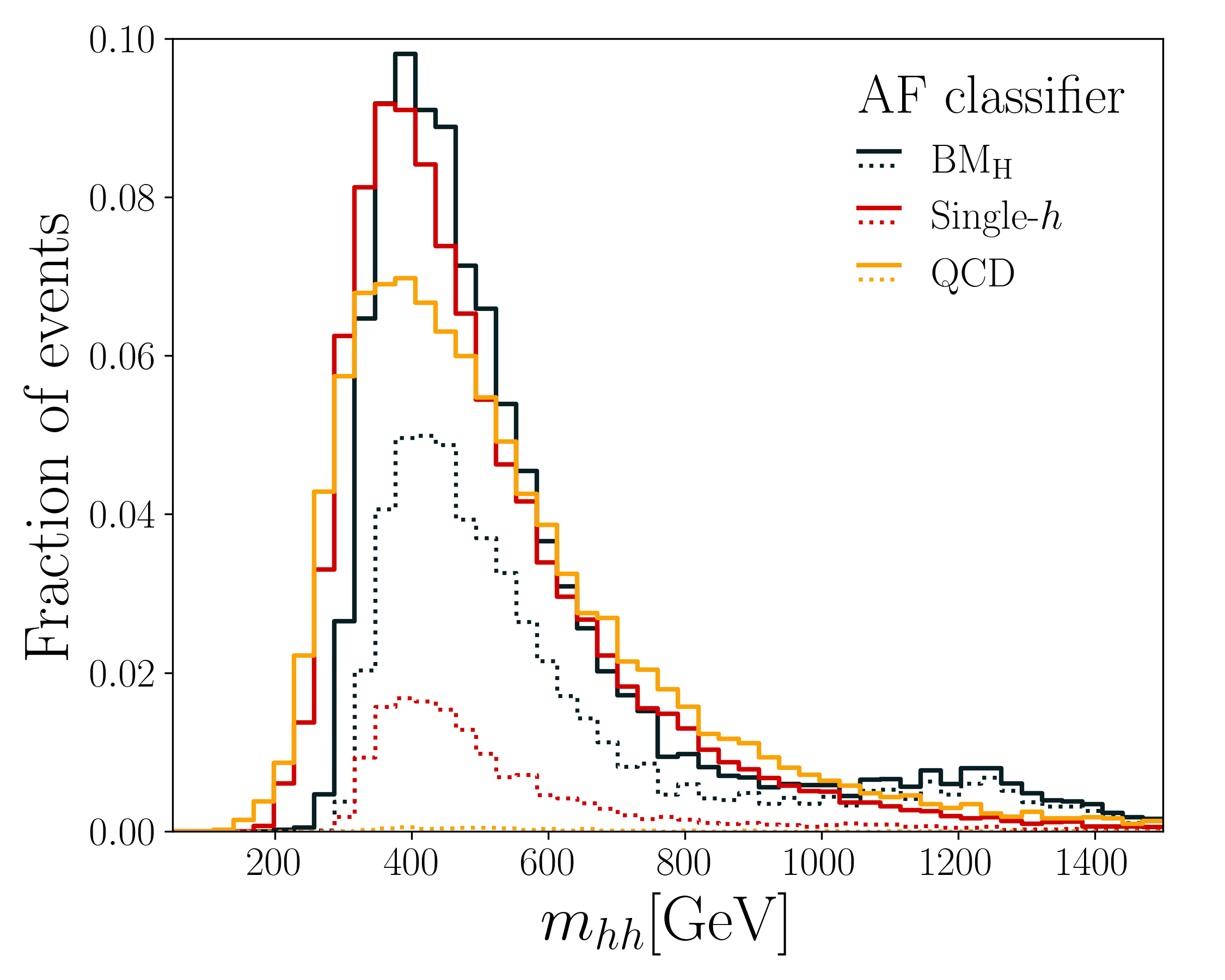}
    \caption{$\Delta R_{\gamma\gamma}$ (left panel) and $m_{hh}$ (right panel) distributions for BM$_{\mathrm{H}}$, before (solid lines) and after (dotted lines) the classification with the AF classifier.}
    \label{fig:distBM3}
\end{figure}

In Figs.~\ref{fig:distBM1} and \ref{fig:distBM3} we display the $\Delta R_{\gamma\gamma}$ and $m_{hh}$ distributions for the benchmarks BM$_{\mathrm{L}}$ and BM$_{\mathrm{H}}$, respectively, before (solid lines) and after (dotted lines) the classification. For BM$_{\mathrm{L}}$, we see from the $m_{hh}$ distribution that almost all of the signal events that are correctly classified populate the second peak at $\simeq 2 \,  m_{\mathrm{NP}}$. The classifier successfully rejects most of the backgrounds events, reaching acceptances below $10^{-2}$ and $10^{-4}$ for the Single-$h$ and QCD contributions, respectively \footnote{As we mentioned before, we also considered $c\bar{c}\gamma\gamma$, however its contribution is of the order of 4\% of $b\bar{b}\gamma\gamma+b\bar{b}\gamma j$ after applying the selection cuts. In addition, its FPR is similar to the QCD backgrounds ($\sim 10^{-4}$). For this reason, we will discard its contribution from now on.}. These small fractions of background events misidentified as signal arise mostly from the region of large $m_{hh}$ values. A similar behavior can be seen from the $\Delta R_{\gamma\gamma}$ distribution, with most of the signal events with photons coming from a boosted Higgs being correctly classified and background events lying in this same region being slightly more difficult to filter. The AF classifier is properly capturing the effects of the lightest new state from the input data. As it is clear from the distributions in Fig.~\ref{fig:distBM3}, these effects are weakened for BM$_{\mathrm{H}}$. In this case,  the AF classifier is correctly classifying most of the signal events in the second peak but, since that region is sparsely populated, a significant fraction of events from the top peak are now kept, at the expense of having more background events misclassified as signal in that region.

For $\mathcal{L}=3$ ab$^{-1}$, we obtain significances of 7.3 and 3.1 for BM$_\mathrm{L}$ and BM$_\mathrm{H}$, respectively. For BM$_\mathrm{H}$, the signal acceptance doubles but the background acceptances increase by approximately two orders of magnitude. This, along with the smaller signal-to-background ratio before the classification, explains the drop in significance with respect to BM$_\mathrm{L}$. On the other hand, for BM$_\mathrm{L}$, a luminosity of $\mathcal{L}=1.4$ ab$^{-1}$ would be enough to reach the discovery level. { These results are subject to changes on the signal cross-section due to uncertainties from top-mass effects, as well as variations on $\alpha_\text{S}$, the renormalization and factorization scale and the PDF set. For the SM, the top mass corrections are the dominant source of uncertainties, reducing the K-factor in a $\sim 15\%$ (see Fig. 18 in Ref. \cite{Baglio:2020ini}). Taking this value as a reference, and reducing the signal rate accordingly, the luminosity for discovery level for BM$_\text{L}$ would increase to $1.8\text{ ab}^{-1}$, while the significances for $\mathcal{L}=3$ ab$^{-1}$ decrease to 6.5 (BM$_\text{L}$) and 2.7 (BM$_\text{H}$).} A rough estimation of the impact of systematic uncertainties on the total background rate $b$ can be obtained by using a modified version of Eq.~(\ref{eq:sig}) provided in Ref.~\cite{Cowan2012DiscoverySF}. With a 15\% of uncertainty, an increase of $5\%$ in luminosity would be required to keep the significance at the discovery level. Moreover, with $\mathcal{L}=3$ ab$^{-1}$, $\mathcal{S}$ stays above 5 for uncertainties as large as 70\%. In the case of BM$_\mathrm{H}$, at this luminosity, a 15\% of uncertainty in the background degrades the significance to 2.7, while for 7\% it is still above the evidence level~\footnote{An estimation of the impact on the significances due to different sources of theoretical uncertainties on the background cross-sections can be found in our previous study~\cite{DaRold:2023hst}, being at most 14\%.}.

Finally, for the most challenging benchmark, BM$_\mathrm{H}$, it is interesting to consider an ideal situation in which only the main backgrounds from QCD are present. We find in that case that the significance increases only to $\sim$ 3.6, which could be interpreted as an upper bound in the attainable significance for our testing sample.\footnote{Notice that there is no clearly identifiable region in the output plane dominated by background events coming exclusively from Single-h.} We therefore conclude that most likely, in order to reach discovery significances, the use of alternative classification algorithms would be necessary. 

\subsection{Analysis with low-level features only}

In this section we study the possibility of maximizing the significance with classifiers built including only low-level features in the input layer. Even though based on previous studies some of the most discriminating high-level features are known for the signal considered, for other processes that may not be the case, and therefore it would be important to explore to what extent this kind of algorithms can overcome the lack of external physical input.

We consider the same architecture from the previous section, modifying the input layer to contain now the 14 low-level variables associated with the momenta of the photons and $b$-jets in the final state (see Table~\ref{tab:vars}). We combine again two classifiers, each one dedicated to separate one of the two main background sources from the signal; the same algorithm used for the AF classifier is applied (see Table~\ref{tab:algo}). Throughout this paper we will denote the resulting classifier as LLF (low-level features) classifier. 

\begin{figure}[h!]
    \centering
    \includegraphics[width=0.5\linewidth]{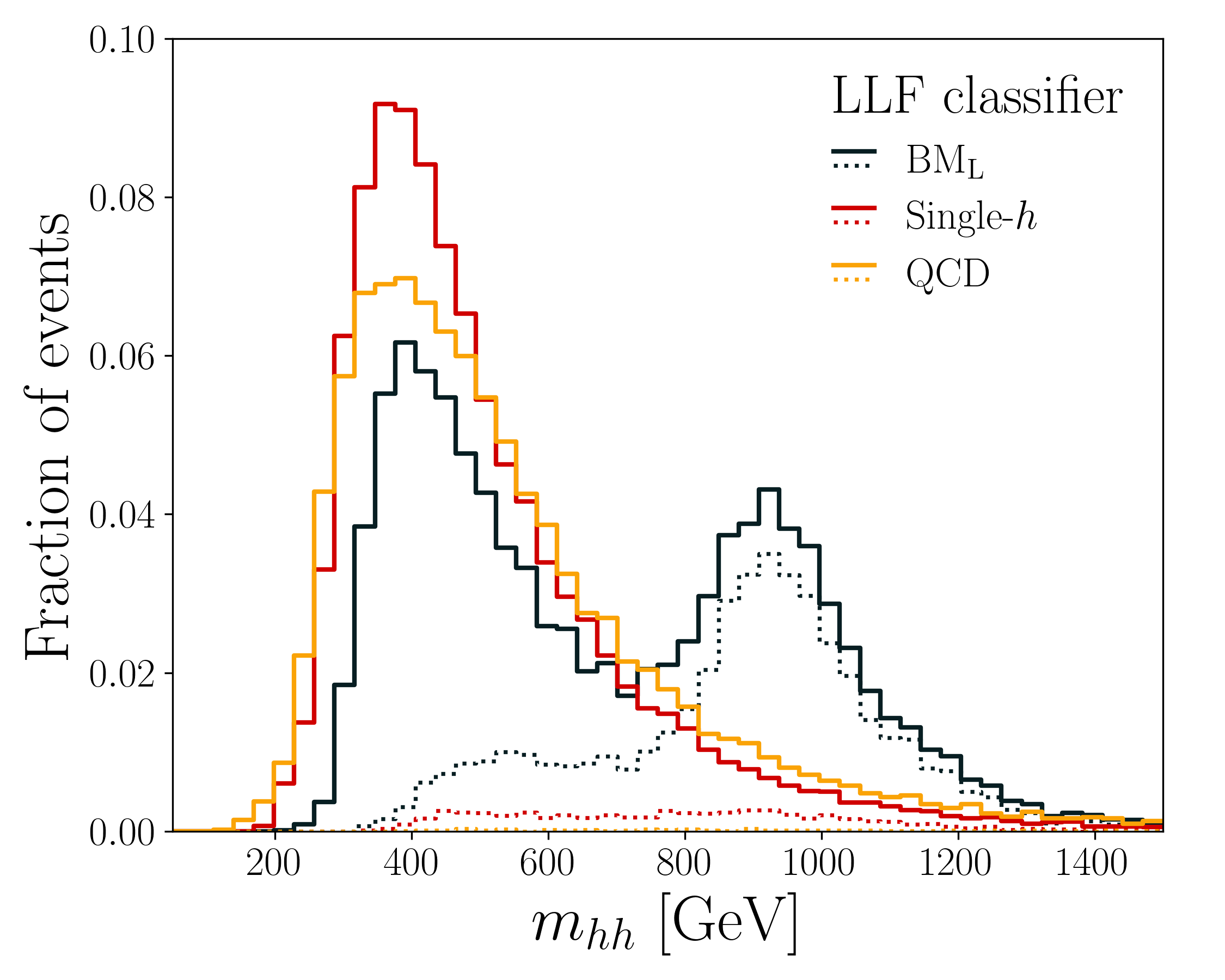}
    \caption{$m_{hh}$ distribution for BM$_{\mathrm{L}}$ before (solid lines) and after (dotted lines) the classification with the LLF classifier.}
    \label{fig:mhh-LLF}
\end{figure}

For BM$_{\mathrm{L}}$, the signal acceptance reached with the LLF classifier increases by $\sim 25$\% with respect to the AF classifier, but the background rejection is significantly worse, resulting in a decrease in the significance, which in this case does not exceed 5.6 for a luminosity of 3~ab$^{-1}$. In Fig.~\ref{fig:mhh-LLF} we show the $m_{hh}$ distribution before and after applying the LLF classifier. Again most of the signal events populating the peak at 2 $m_{\mathrm{NP}}$ are correctly classified, but in contrast to the AF classifier (see Fig.~\ref{fig:distBM1}), more signal events lying at smaller $m_{hh}$ values are kept. This increase in accepted signal events comes at the expense of more background events being misidentified, as it is clear from the $m_{hh}$ distribution for Single-$h$. Even when it is not visible in Fig.~\ref{fig:mhh-LLF}, the acceptance for the QCD background increases by approximately one order of magnitude for the LLF classifier. 

\begin{figure}[h!]
    \centering
    \includegraphics[width=0.5\linewidth]{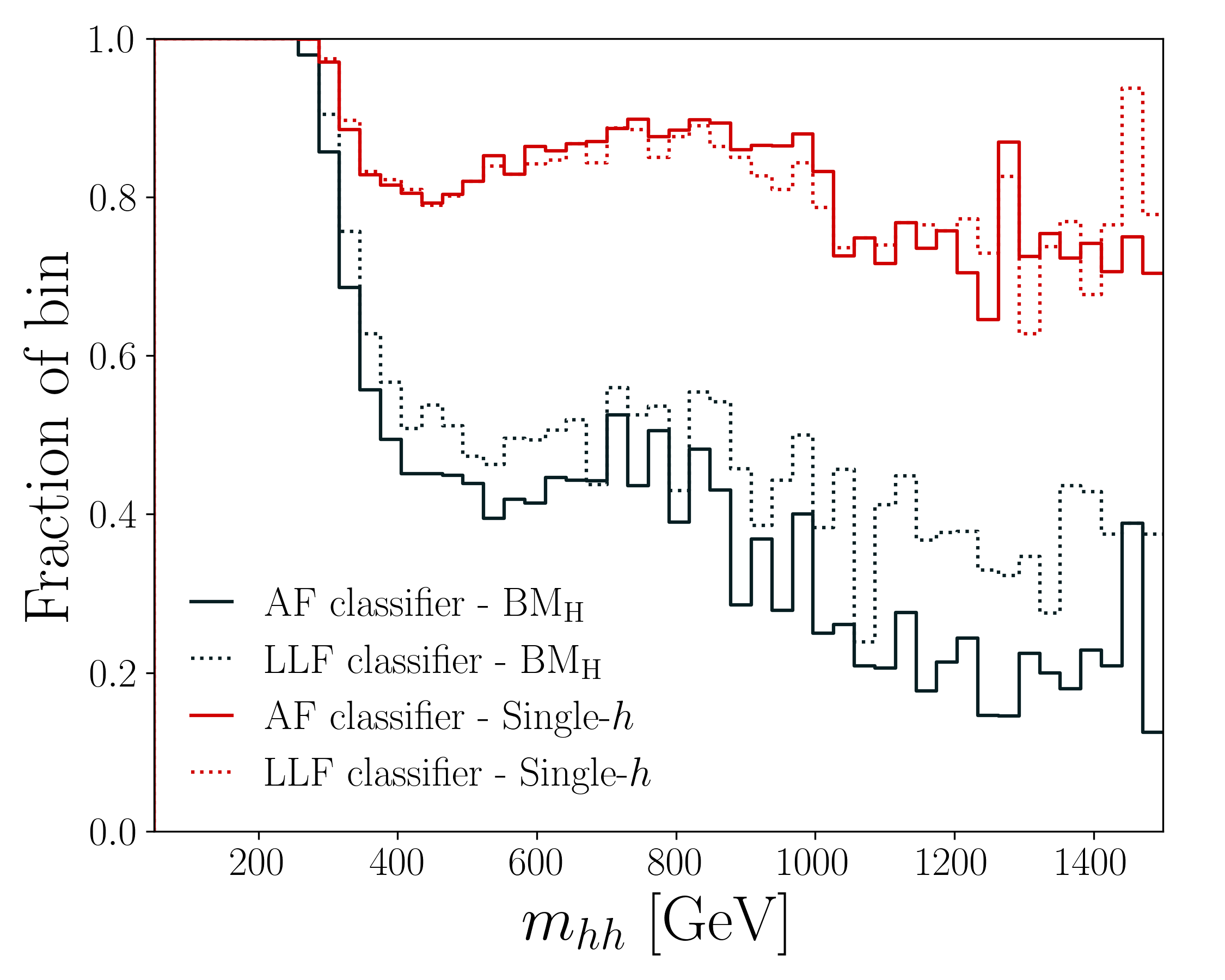}
    \caption{Fraction of bins in $m_{hh}$ rejected by the AF (solid lines) and LLF (dotted lines) classifiers for BM$_{\mathrm{H}}$ (black) and Single-$h$ background (red). QCD backgrounds are not shown because they would not be visible in the figure (their acceptance being of order $10^{-4}$).}
    \label{fig:excl-bin-mhh}
\end{figure}

For BM$_{\mathrm{H}}$, the performance of the classification of the signal and the backgrounds events worsen relative to the AF classifier, which leads to a drop in the significance below the evidence level, specifically yielding 2.4 for $\mathcal{L}=3$ ab$^{-1}$. In Fig.~\ref{fig:excl-bin-mhh} we compare the AF and LLF classifiers in terms of the fraction of each bin of the $m_{hh}$ distribution that is rejected. We see that the signal is better identified by the AF classifier, specially in the region above 800 GeV, and also the Single-$h$ background is slightly better rejected. Furthermore, the QCD background rejection from the LLF classifier deteriorates considerably, resulting in a 45\% increase in its acceptance that has a major impact on the significance drop.

\begin{figure}[h!]
    \centering
    \includegraphics[width=0.45\linewidth]{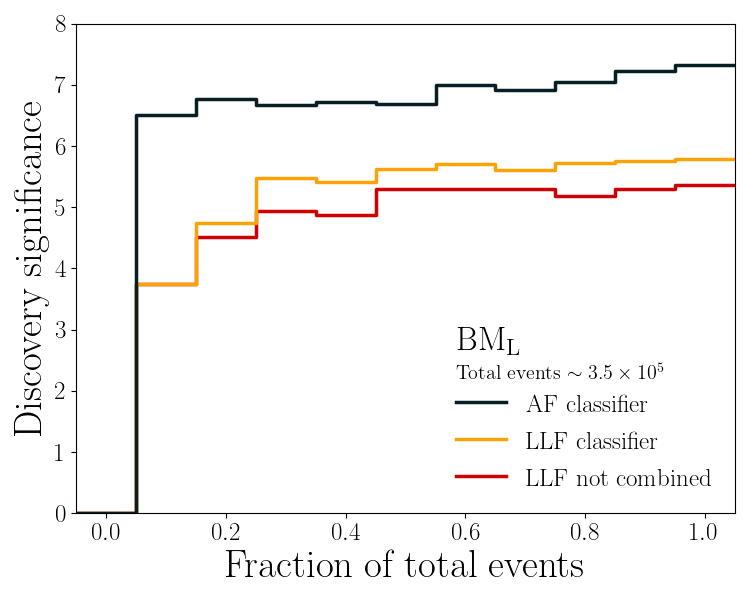}
    \includegraphics[width=0.45\linewidth]{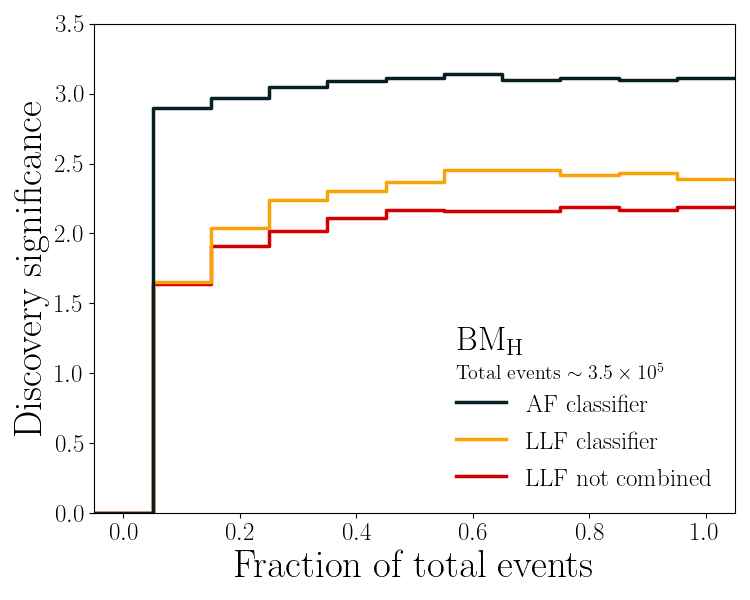}
    \caption{Evolution of the discovery significance with the fraction of total events included during the training for the AF (black) and LLF (yellow) classifiers. An alternative low-level features classifier trained including events of both backgrounds is also displayed in red.}
    \label{fig:significance}
\end{figure}

It is interesting to study the evolution with the fraction of total events included in the training set of the maximum significance obtained with the AF and LLF classifiers. In Fig.~\ref{fig:significance} we show the discovery significance obtained when the fraction of total events used during the training is increased from 10\% to 100\%. In addition to the AF (black line) and LLF (yellow line) classifiers, we also include an alternative low-level classifier which is trained with both backgrounds at once (red line). For both benchmarks the significance for the AF classifier saturates relatively fast, surpassing the discovery (evidence) level for $\sim 10$\% ($\sim 40$\%) of the total events for BM$_{\mathrm{L}}$ (BM$_{\mathrm{H}}$). The AF classifier outperforms the LLF regardless the fraction of events, but the gap between them decreases with the number of events in the training set until the fraction of events reaches the $\sim 50$\% for BM$_{\mathrm{L}}$ ($\sim 60$\% for BM$_{\mathrm{H}}$). Note that the significance corresponding to the LLF classifier is systematically above that obtained with the alternative low-level classifier, which shows the advantage of combining classifiers trained exclusively with each background, QCD and Single-$h$.

\begin{figure}[h!]
    \centering
    \includegraphics[width=0.8\linewidth]{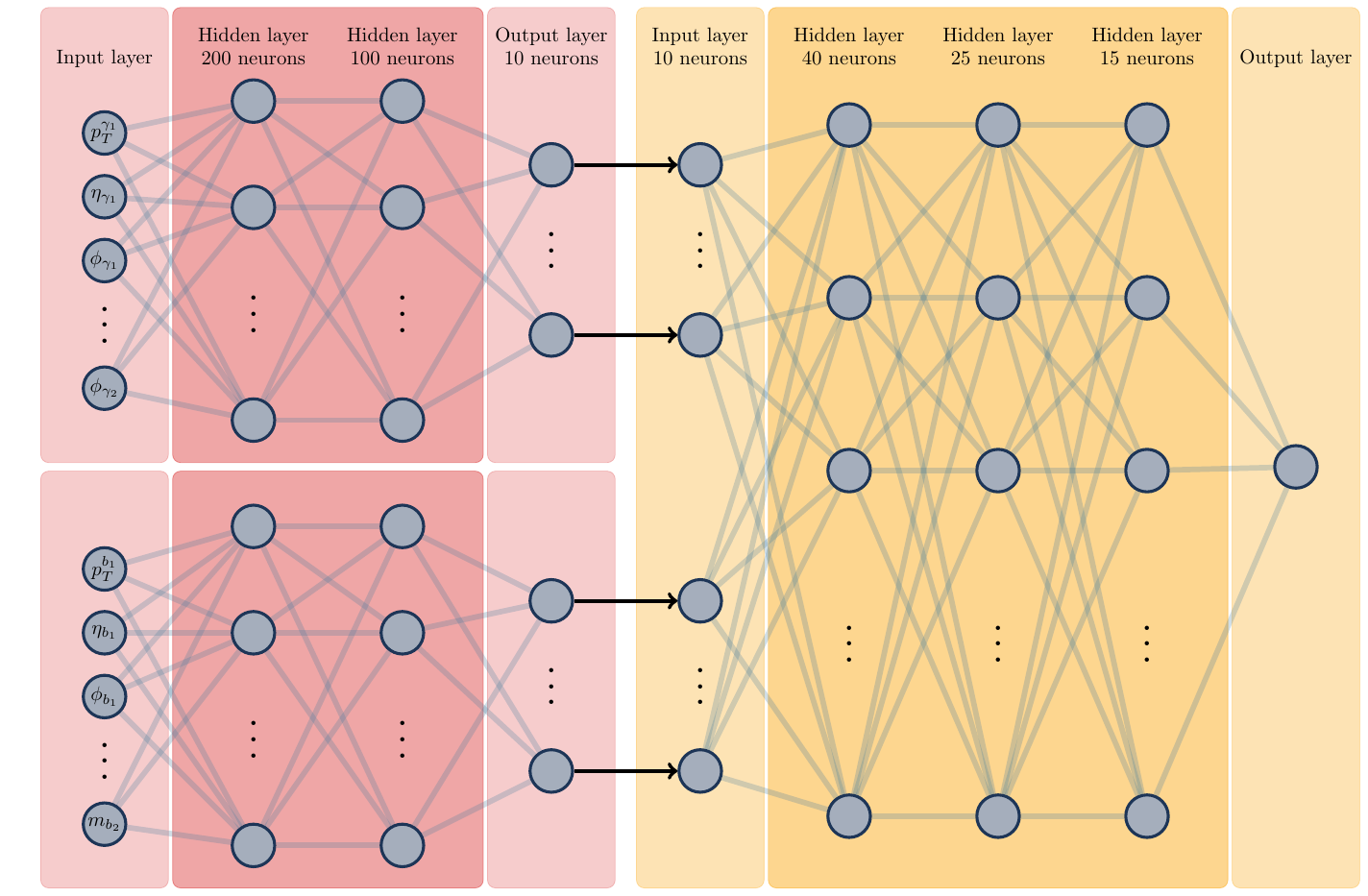}
    \caption{Alternative DNN architecture tested in BM$_{\mathrm{H}}$. Two sub-networks are fed with the momenta of the $b$-jets and the photons, respectively, and their outputs feed a third network from which the final output is obtained.}
    \label{fig:arch-subn}
\end{figure}

\begin{figure}[h!]
    \centering
    \includegraphics[width=0.45\linewidth]{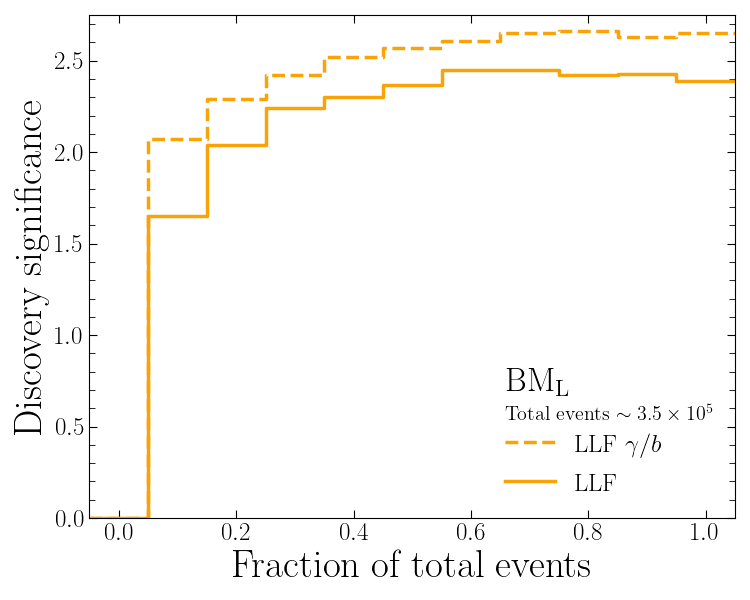}
    \caption{Evolution of the discovery significance with the fraction of total events included to train the classifiers. We show the LLF classifier (solid line) along with an alternative low-level features classifier (LLF $\gamma/b$) in which the momenta of the $b$-jets and the photons are pre-processed in sub-networks (dashed line).}
    \label{fig:significance2}
\end{figure}

We conclude this section by presenting a low-level classifier with an alternative architecture for the most challenging benchmark, BM$_{\mathrm{H}}$. The new structure is depicted in Fig.~\ref{fig:arch-subn} and  incorporates two sub-networks, one fed by the momenta of the photons and the other by the momenta of the $b$-jets. With this architecture we attempt to compensate the lack of the most relevant high-level features associated to one of the Higgs bosons, such as $m_{\gamma\gamma}$ or $m_{bb}$, with the pre-processing of low-level variables in the sub-networks. We again combine two classifiers following the algorithm described in Table~\ref{tab:algo}. The resulting classifier is denoted as LLF $\gamma/b$. In Fig.~\ref{fig:significance2} we compare the significance corresponding to the classifiers LLF and LLF $\gamma/b$ in terms of the fraction of total events included in the training set. We see that the LLF $\gamma/b$ classifier outperforms the LLF for all the fraction of events, although the maximum significance achieved is 2.7, which corresponds to the complete training sample being used and entails a slight improvement of 10\% with respect to the LLF classifier. This improvement is not enough to match the performance of the AF classifier, which seems to be unlikely to accomplish with classifiers based on low-level features only and MLP networks.

\section{Conclusions}
\label{sec:conclusions}
The measurement of di-Higgs production represents a central goal in the exploration of the Higgs potential and the mechanism of electroweak symmetry breaking. In the Standard Model, the smallness of the production cross-section makes this channel experimentally challenging, motivating the study of scenarios beyond the SM that could enhance the signal. Among these, models with new colored scalar states—such as scalar leptoquarks or squark-like partners—provide a minimal and theoretically well-motivated framework to increase the di-Higgs yield. In this work we have explored the power of machine learning techniques, specifically multiple layer perceptron neural networks, for improving the search for non-resonant di-Higgs production in the $b\bar{b}\gamma\gamma$ final state at the HL-LHC. Within the framework of a BSM scenario with new colored scalars, we examined two representative benchmark points, BM$_{\rm L}$ and BM$_{\rm H}$, consistent with current collider and flavor constraints, with lightest new state masses of 464 GeV and 621 GeV, respectively.

We have implemented a complete simulation framework incorporating one-loop level contributions of colored scalar fields and the main Standard Model backgrounds. To maximize the power to discriminate signal from background, we explored a set of DNN classifiers with different input layers and architectures. A key finding from our analysis is the superior performance achieved by employing two dedicated DNN classifiers—one targeting QCD-induced backgrounds and another focusing on single-Higgs backgrounds—compared to a single classifier handling all backgrounds simultaneously. This approach proved particularly effective in leveraging the distinctive kinematic features of each background type. While the invariant masses $m_{\gamma\gamma}$ and $m_{bb}$ emerged as the most powerful discriminating variables, the inclusion of additional kinematic observables further strengthened the classification performance.

For the All Features (AF) classifier, which incorporates both low- and high-level variables, we obtained promising results: significances well above the discovery level (7.2) for BM$_{\rm L}$ and at the evidence level (3.1) for BM$_{\rm H}$, with an integrated luminosity of $3\ \text{ab}^{-1}$. In fact, for BM$_{\rm L}$, discovery level can be obtained for $1.4\ \text{ab}^{-1}$. If the signal rate is reduced in a $\sim 15\%$, accounting for possible theoretical uncertainties in our estimation of the signal cross-section, the significances decrease to 6.5 (BM$_\text{L}$) and 2.7 (BM$_\text{H}$), and a luminosity of $1.8\text{ ab}^{-1}$ would be required to reach the discovery level for $\text{BM}_\text{L}$. The AF classifier showed particular efficiency in identifying signal events within the distinctive region of the $m_{hh}$ distribution around 2 $m_{\text{NP}}$, where BSM contributions are most pronounced.

The comparison with classifiers using only low-level features (LLF) revealed the crucial importance of high-level variables. The significant performance gap, especially evident for the more challenging BM$_{\rm H}$ scenario where the LLF significance dropped to 2.4, underscores the value of physically meaningful constructed variables. While an alternative architecture with dedicated sub-networks for photons and $b$-jets provided modest improvement, it could not match the performance of the AF classifier.

Note that the significances obtained with neural networks are better than those from our previous article \cite{DaRold:2023hst}, based on rectangular cuts and kinematical distributions. In that case, we obtained 5.3 for BM$_{\rm L}$ and 2.3 for BM$_{\rm H}$, with an integrated luminosity of $3\ \text{ab}^{-1}$. 

Looking forward, these results highlight the substantial potential of machine learning approaches to maximize the physics reach of the HL-LHC. For particularly challenging scenarios like BM$_{\rm H}$, where the significance remains below discovery threshold, alternative classification algorithms or additional discriminating variables may be necessary to achieve discovery level. We plan to generalize this study beyond the two benchmarks chosen in a future work.

\section*{Acknowledgments}
This work has been partially supported by CONICET and ANPCyT
projects PIP-11220200101426 and PICT-2018-03682. The authors acknowledge the use of the HPC cluster of the Physics Department at Centro Atómico Bariloche (CNEA), which is part of Argentina’s National High-Performance Computing System (SNCAD) and the IFLP Physics Cluster for the computational resources provided, as well as F. Alonso, G. Berman, G. Mu\~noz and H. Wahlberg for technical support. They also wish to thank A. Perez for a careful reading of the article and M. Szewc for useful discussions. LD acknowledges the hospitality of IFLP during the completion of this work.  AM acknowledges the hospitality of Perimeter Institute during the completion of this work. This research was supported in part by Perimeter Institute for Theoretical Physics. Research at Perimeter Institute is supported by the Government of Canada through the Department of Innovation, Science, and Economic Development, and by the Province of Ontario through the Ministry of Colleges and Universities.

\bibliography{biblio}
\bibliographystyle{JHEP}

\end{document}